\documentstyle[multicol,aps,psfig]{revtex}
\renewcommand{\narrowtext}{\begin{multicols}{2}
\global\columnwidth20.5pc\noindent}
\renewcommand{\widetext}{\end{multicols}
\global\columnwidth42.5pc}
\begin{document}
\draft
\preprint{28 February 2002}
\title{Fermionic description of spin-gap states of antiferromagnetic
       Heisenberg ladders\\
       in a magnetic field}
\author{Hiromitsu Hori and Shoji Yamamoto}
\address
{Department of Physics, Okayama University,
 Tsushima, Okayama 700-8530, Japan}

%\date{Received \hspace{6cm}}
\date{Received 28 February 2002}
\maketitle
%\begin{abstract}
%\end{abstract}
\narrowtext

   Spin gaps$-$the energy gaps in magnetic excitation spectra$-$of
low-dimensional Heisenberg antiferromagnets have been attracting
considerable interest in recent years.
This fascinating subject was initiated by Haldane \cite{H1153},
renewed at the opportunity of high-temperature superconductivity
being discovered \cite{B188}, and further developed via the
synthesis of ladder materials \cite{J219,H230}.
Dagotto, Riera and Scalapino \cite{D5744} pioneeringly pointed out a
possible electronic mechanism for the formation of spin gaps in
two-leg ladders.
Following investigations \cite{D618} revealed that assembling chains
into ladders, the crossover between one and two dimensions is far
from smooth$-$Heisenberg ladders with an even number of legs have a
spin gap, while those with an odd number of legs have no gaps.

   Spin ladders in a magnetic field provide further interesting
topics.
Their ground-state magnetization curves have extensively been
studied \cite{C5126,C6241,C1768,T396} in an attempt to find {\it
quantized plateaux}.
Spin-$S$ $L$-leg Heisenberg ladders
\begin{equation}
   {\cal H}
   =\sum_{n=1}^N
    \left(
     J\sum_{l=1}^{L}
      \mbox{\boldmath$S$}_{n,l}\cdot\mbox{\boldmath$S$}_{n+1,l}
    +J'\sum_{l=1}^{L-1}
      \mbox{\boldmath$S$}_{n,l}\cdot\mbox{\boldmath$S$}_{n,l+1}
    \right)\,,
   \label{E:H}
\end{equation}
may exhibit magnetization plateaux at $M/M_{\rm sat}=m/LS$ provided
$LS-m\in\mbox{\boldmath Z}$ \cite{C5126,O1984,T103}, where $M$ and
$M_{\rm sat}$ are the magnetization and its saturated value,
respectively.
Since coupled-chain materials are likely to be systematically
obtained in the case of spin $\frac{1}{2}$ \cite{A3463}, most of
numerical efforts are being devoted to $S=\frac{1}{2}$ ladders.
Then, without any bond polymerization, the most tractable system of
our interest is the three-leg ladder.
Lanczos diagonalization of finite systems \cite{C5126}, a
series-expansion technique \cite{C6241} and density-matrix
renormalization-group calculations \cite{T396} are all in agreement
to support the existence of the plateau at $M/M_{\rm sat}=1/3$ for
strong interchain coupling $J'\agt J$.
As for more-than-three-leg ladders, the existence of plateaux as well
as their surviving region is still left to be verified.

   Spin-$\frac{1}{2}$ two- and three-leg ladders are indeed realized
in layer materials Sr$_{n-1}$Cu$_{n+1}$O$_{2n}$ ($n=3,5,\cdots$)
\cite{H230,A3463}, while four- and five-leg ones in
La$_{4+4n}$Cu$_{8+2n}$O$_{14+8n}$ ($n=2,3,\cdots$) \cite{B}.
It is unfortunate that nevertheless large degrees of freedom prevent
us from making numerical access to multi-leg ladders.
In such circumstances, we propose a systematic approach to Heisenberg
ladders in order to describe their spin-gap states in a field.
Employing the Jordan-Wigner transformation on a unique path and then
making a mean-field treatment of the fermionic Hamiltonian, we
semiquantitatively visualize the appearance of plateaux and estimate
the corresponding critical interactions $J'_{\rm c}$ at an
arbitrary number of legs, $L$.

   It is along a snake-like path,
$(n,l)
=(1,1)\rightarrow(1,2)\rightarrow\cdots\rightarrow(1,L)\rightarrow
 (2,L)\rightarrow(2,L-1)\rightarrow\cdots\rightarrow(2,1)\rightarrow
 (3,1)\rightarrow\cdots$,
that we define spinless fermions.
This elaborately ordered path was first proposed by Dai and Su
\cite{D964} and turned out to describe the spin gap as a function of
$L$ much better than the naively ordered path \cite{A6233}, which is
usually employed.
When we introduce renumbered spin operators
$\widetilde{\mbox{\boldmath$S$}}_{n,l}=\mbox{\boldmath$S$}_{n,l}$
($\mbox{\boldmath$S$}_{n,L-l+1}$) for an odd (even) $n$, the spinless
fermions are created as
$c_{n,l}^\dagger=\widetilde{S}_{n,l}^+
 {\rm exp}[-{\rm i}\pi
 (\sum_{i=1}^{n-1}\sum_{j=1}^L
   \widetilde{S}_{i,j}^+\widetilde{S}_{i,j}^-
 +\sum_{j=1}^{l-1}
   \widetilde{S}_{n,j}^+\widetilde{S}_{n,j}^-)]$.
Making a standard mean-field treatment \cite{D964} of the fermionic
Hamiltonian and assuming the spatial homogeneity
$\langle\widetilde{S}_{n,l}^z\rangle
 =\langle c_{n,l}^\dagger c_{n,l}\rangle-1/2=M/LN$, we obtain
\begin{eqnarray}
   &&
   {\cal H}
   =\frac{M}{LN}\sum_k\sum_{l=1}^L
    \Bigl[
     2J+(2-\delta_{l,1}-\delta_{l,L})J'
    \Bigr]
    c_{k,l}^\dagger c_{k,l}
   \nonumber \\
   &&\qquad
   +\frac{J'}{2}\sum_k\sum_{l=1}^{L-1}
    \left(
     c_{k,l}^\dagger c_{k,l+1}+c_{k,l+1}^\dagger c_{k,l}
    \right)
   \nonumber \\
   &&\qquad
   +\frac{J}{2}\sum_k\sum_{l=1}^L
    \Bigl[
     {\rm e}^{-{\rm i}\pi(L-l)(2M/LN+1)}
     {\rm e}^{-{\rm i}k}
   \nonumber \\
   &&\qquad\quad
    +{\rm e}^{ {\rm i}\pi(l-1)(2M/LN+1)}
     {\rm e}^{ {\rm i}k}
    \Bigr]
    c_{k,l}^\dagger c_{k,L-l+1}
   \nonumber \\
   &&\qquad
   -\Bigl(J+\frac{L-1}{L}J'\Bigr)M\Bigl(1+\frac{M}{LN}\Bigr)\,,
   \label{E:MFH}
\end{eqnarray}
where
$c_{k,l}^\dagger=N^{-1/2}\sum_n{\rm e}^{-{\rm i}kn}c_{n,l}^\dagger$.
The effective Hamiltonian (\ref{E:MFH}), together with the Zeeman
term $-H\sum_k\sum_{l=1}^L(c_{k,l}^\dagger c_{k,l}-1/2)$, is
numerically diagonalized, adopting the open boundary condition along
rungs, while taking the thermodynamic limit under the periodic
boundary condition along legs.

\begin{figure}
\centerline
{\mbox{\psfig{figure=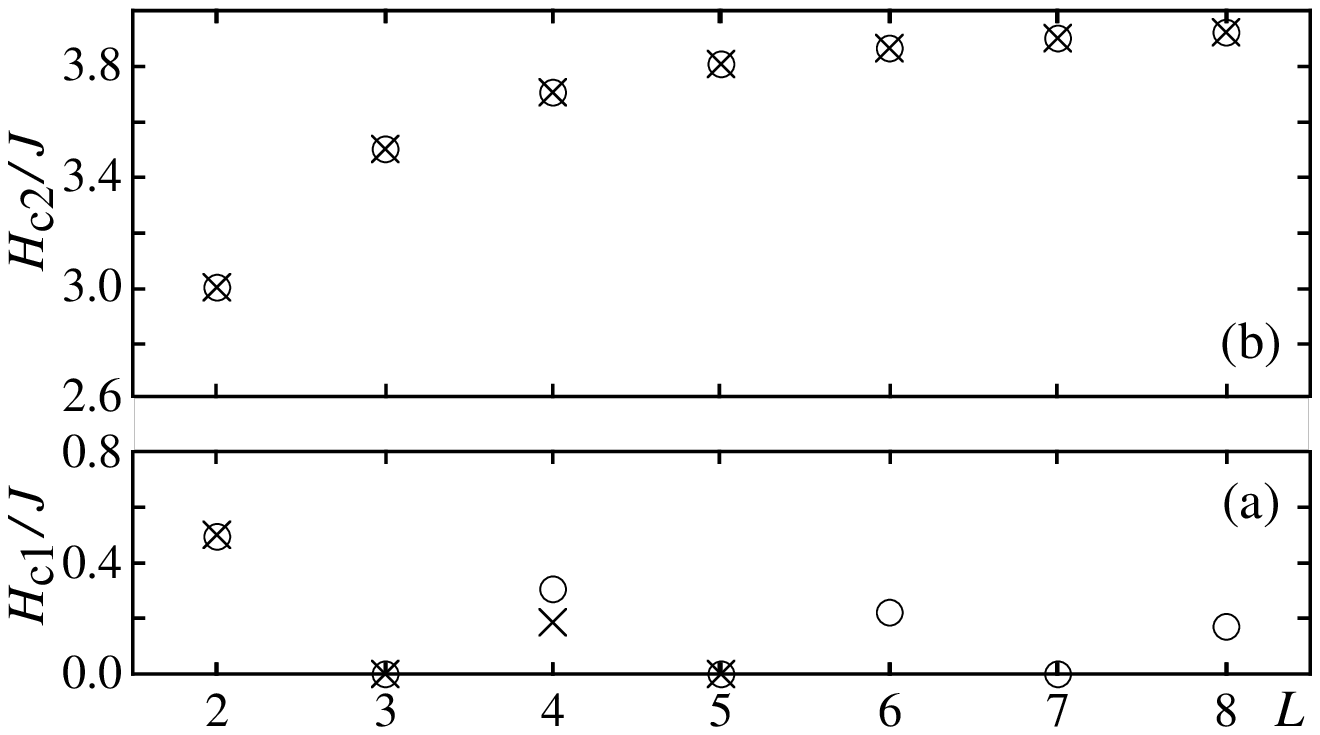,width=99mm,angle=0}}}
\vspace*{2mm}
\caption{Spin gaps $H_{\rm c1}$ (a) and saturation fields
         $H_{\rm c2}$ (b) for the $L$-leg ladders with $J'=J$, where
         the present culculations ($\circ$) are compared with
         density-matrix renormalization-group estimates [17]
         ($\times$) in (a) and with the exact values ($\times$) in
         (b).}
\label{F:Hc}
\vspace*{-36mm}
\centerline
{\mbox{\psfig{figure=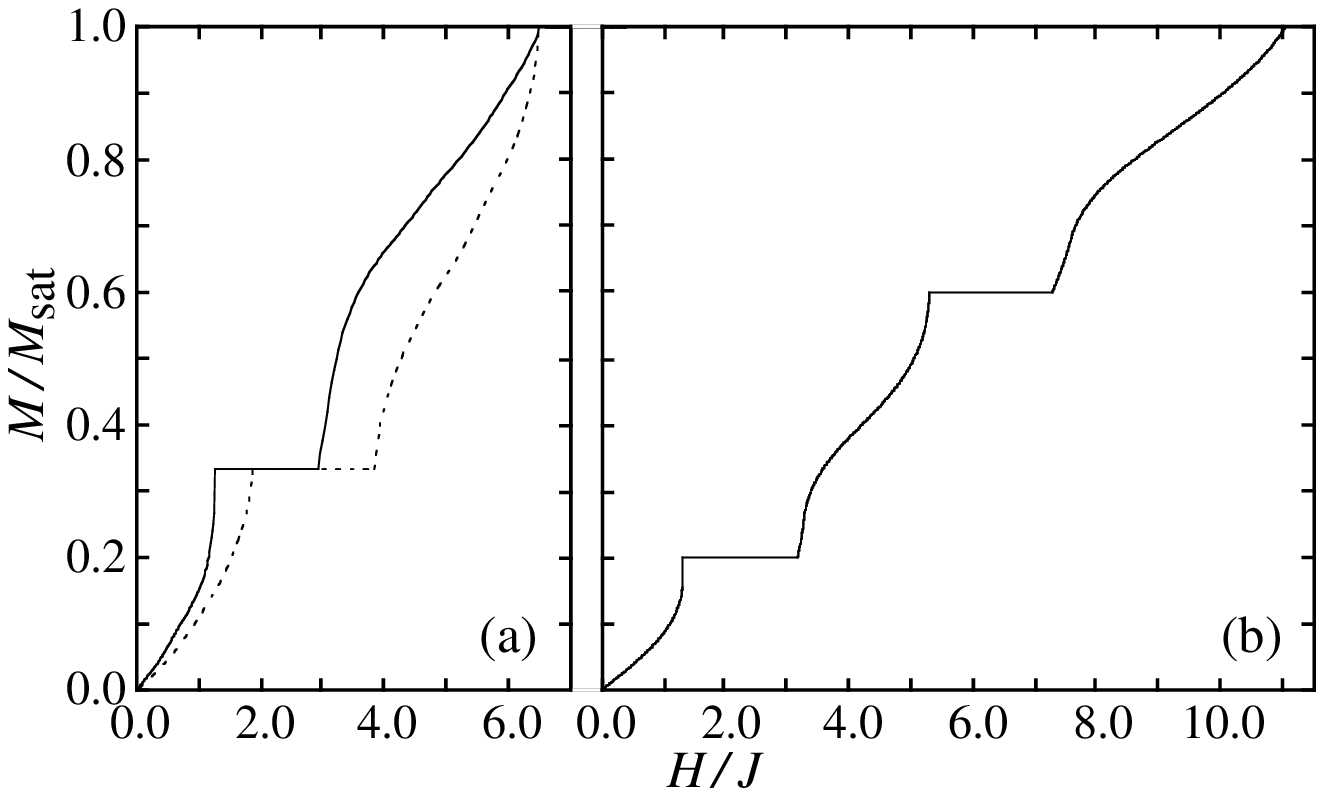,width=99mm,angle=0}}}
\vspace*{2mm}
\caption{Ground-state magnetization curves of the three- and five-leg
         ladders with $J'=3J$ (a) and $J'=5J$ (b), respectively.
         Numerical-diagonalization results [7] are also shown by
         a dotted line in (a).}
\label{F:M}
\end{figure}

   In order to verify the reliability of the present approach, we
plot in Fig. \ref{F:Hc} the thus-obtained spin gaps ($H_{{\rm c}1}$)
and saturation fields ($H_{{\rm c}2}$), the former of which are
compared with highly precise numerical estimates \cite{W886}, whereas
the latter of which with the exact solutions
$H_{{\rm c}2}=2J+J'[1+\cos(\pi/L)]$.
Although the present scheme somewhat overestimates $H_{{\rm c}1}$
with increasing $L$, it correctly tells whether the gap survives or
not.
As for $H_{{\rm c}2}$, the present calculation can be regarded as
exact.
We further show in Fig. \ref{F:M} typical calculations of the
ground-state magnetization.
All the plateaux satisfying the criterion
$L/2-M/N\in\mbox{\boldmath Z}$ indeed appear with
increasing $J'$.
Mean-field approaches generally underestimate quantum fluctuations
and therefore necessarily overestimate the magnetization.
Allow us, however, to stress that the length of a plateau is still
well describable in our treatment, which is essential to estimate
the lower boundaries of the plateau-surviving region.
\vspace*{-40mm}

\begin{figure}
\centerline
{\mbox{\psfig{figure=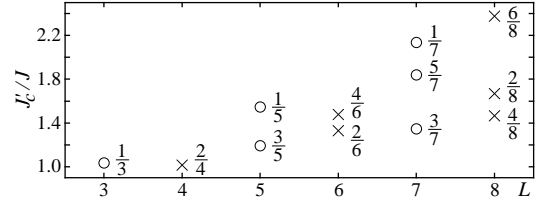,width=99mm,angle=0}}}
\vspace*{2mm}
\caption{Critical interchain interaction to induce magnetization
         plateaux, $J'_{\rm c}$, for the $L$-leg ladders, where the
         corresponding magnetization values $M/M_{\rm sat}$ are
         indicated beside symbols.}
\label{F:J'c}
\end{figure}

   Now we explore the main issue in Fig. \ref{F:J'c}.
The critical value $J'_{\rm c}=1.04\pm 0.01$ for $L=3$ is in good
agreement with the previous estimate $J'_{\rm c}\simeq J$
\cite{C6241} obtained through the series expansion from the
strong-rung-coupling limit $J/J'\rightarrow 0$.
$J'_{\rm c}$ appears to increase with $L$, where even- and odd-$L$
ladders may form distinguishable series of their own.
Here is a conclusive remark:
Among possible nontrivial quantized magnetizations,
$M/M_{\rm sat}=1-2/L,1-4/L,\cdots,2/L\,(1/L)$,
{\it the inner plateaux are easier to induce}.
The plateaux at the end-value magnetizations
$M/M_{\rm sat}=1-2/L,2/L\,(1/L)$ are generally induced with larger
$J'$ than those at the inner-value magnetizations are.
Practical observation of multi-plateau magnetization curves may
less be feasible with increasing $L$.
At the least, however, an available five-leg ladder material
(La$_2$CuO$_4$)$_3$La$_2$Cu$_4$O$_7$ \cite{B} encourages us
to make theoretical explorations into the unique way from one- to
two-dimensional antiferromagnets in a field.
It is also important, further numerical verification of the plateau
at $M/M_{\rm sat}=3/5$ more surviving than that at
$M/M_{\rm sat}=1/5$ for $L=5$.
\vspace*{-46mm}

   We are grateful to Prof. M. Takahasi for useful comments.
This work was supported by the Japanese Ministry of Education,
Science, Sports and Culture, and by the Sumitomo Foundation.

\widetext

\begin{references}

\bibitem{H1153}
   F. D. M. Haldane:
      Phys. Rev. Lett. {\bf 50} (1983) 1153.

\bibitem{B188}
   J. G. Bednorz and K. A. M\"uller:
      Z. Phys. B {\bf 64} (1986) 188.

\bibitem{J219}
   D. C. Johnston, J. W. Johnson, D. P. Goshorn and A. J. Jacobson:
      Phys. Rev. B {\bf 35} (1987) 219.

\bibitem{H230}
   Z. Hiroi, M. Azuma, M. Takano and Y. Bando:
      J. Solid State Chem. {\bf 95} (1991) 230.

\bibitem{D5744}
   E. Dagotto, J. Riera and D. Scalapino:
      Phys. Rev. B {\bf 45} (1992) R5744.

\bibitem{D618}
   E. Dagotto and T. M. Rice:
      Science {\bf 271} (1996) 618.

\bibitem{C5126}
   D. C. Cabra, A. Honecker and P. Pujol:
      Phys. Rev. Lett. {\bf 79} (1997) 5126.

\bibitem{C6241}
   D. C. Cabra, A. Honecker and P. Pujol:
      Phys. Rev. B {\bf 58} (1998) 6241.

\bibitem{C1768}
   D. C. Cabra and M. D. Grynberg:
      Phys. Rev. Lett. {\bf 82} (1999) 1768.

\bibitem{T396}
   K. Tandon, S. Lal, S. K. Pati, S. Ramasesha and D. Sen:
      Phys. Rev. B {\bf 59} (1999) 396.

\bibitem{O1984}
   M. Oshikawa, M. Yamanaka and I. Affleck:
      Phys. Rev. Lett. {\bf 78} (1997) 1984.

\bibitem{T103}
   K. Totsuka:
      Phys. Lett. A {\bf 228} (1997) 103.

\bibitem{A3463}
   M. Azuma, Z. Hiroi, M. Takano, K. Ishida and Y. Kitaoka:
      Phys. Rev. Lett. {\bf 73} (1994) 3463.

\bibitem{B}
   B. Batlog and R. Cava,
      unpublished.

\bibitem{D964}
   X. Dai and Z. Su:
      Phys. Rev. B {\bf 57} (1998) 964.

\bibitem{A6233}
   M. Azzouz, L. Chen and S. Moukouri:
      Phys. Rev. B {\bf 50} (1994) 6233.

\bibitem{W886}
   S. R. White, R. M. Noack and D. J. Scalapino:
      Phys. Rev. Lett. {\bf 73} (1994) 886.

\end{references}
\end{document}